\documentclass[
    ,final            
  ]
  {aipproc}

\layoutstyle{6x9}

\begin{document}

\title{Gamma-Ray Bursts Observed with the Spectrometer SPI Onboard INTEGRAL}

\author{A. von Kienlin}{
  address={Max-Planck-Institut f\"{u}r extraterrestrische Physik, Giessenbachstrasse, 85748 Garching, Germany}
}

\author{A. Rau}{
  address={Max-Planck-Institut f\"{u}r extraterrestrische Physik, Giessenbachstrasse, 85748 Garching, Germany}
}

\author{V. Beckmann}{
 address={NASA Goddard Space Flight Center, Greenbelt, Maryland 20771, USA}
}
\author{S. Deluit}{
  address={INTEGRAL Science Data Centre, Chemin d' \'Ecogia 16, 1290 Versoix, Switzerland}
}

\begin{abstract}
The spectrometer SPI is one of the main detectors of ESA's INTEGRAL mission. The instrument offers two interesting 
and valuable capabilities for the detection of the prompt emission of Gamma-ray bursts. Within a field of view of 
16\,degrees, SPI is able to localize Gamma-ray bursts with an accuracy of 10\,arcmin. The large anticoincidence 
shield, ACS, consisting of 512\,kg of BGO crystals, detects Gamma-ray bursts quasi omnidirectionally above 
$\sim$70\,keV. Burst alerts from SPI/ACS are distributed to the interested community via the INTEGRAL Burst Alert 
System. The ACS data have been implemented into the 3rd Interplanetary Network and have proven valuable for the 
localization of bursts using the triangulation method. During the first 8\,months of the mission approximately one 
Gamma-ray burst per month was localized within the field of fiew of SPI and 145 Gamma-ray burst candidates were 
detected by the ACS from which 40\,\% have been confirmed by other instruments.

\end{abstract}

\maketitle


\subsection{GRBs observed with the camera of SPI}

The aim of the spectrometer onboard INTEGRAL (SPI) \cite{2003A&A...411L..63V} is to perform high-resolution
spectroscopy of astrophysical sources in the energy range between
20\,keV and 8\,MeV. The imaging capability is good,
but is exceeded by that of the imager IBIS  which complements SPI
by having higher imaging resolution, but lower spectroscopic
resolving power. The detection and investigation of cosmic gamma-ray bursts
(GRBs) is one of the important scientific topics of the INTEGRAL
mission.
%
%
The broad energy coverage of SPI is well suited to constrain the spectral shape,
both below and above the energy at which the GRB power output is typically
peaked ($\sim 250$\,keV) \cite{2000ApJS..126...19P}. The long-standing controversy over the
existence of short-lived spectral features in GRB spectra can
be addressed by SPI's superb spectroscopic capabilities.
In addition, the capability to cross-calibrate both spectra and
images  between the two experiments is extremely important,
particularly in the case of such short-lived events as GRBs, which
cannot be re-observed. Currently GRBs which occur inside SPI's field of view (FoV)
are detected and analysed offline.

Since the start of the mission, six GRBs have been observed within the FoV of IBIS \& SPI. The obtained scientific 
results are presented in
\citet{2003A&A...411L.307M} for GRB021125, \citet{2003A&A...411L.311M} for GRB 021219,
\citet{2003A&A...409..831G} for GRB030131, \citet{2003ApJ...590L..73M} for GRB0302227,
\citet{2003A&A...411L.321V} for GRB030320, and \citet{2003A&A...411L.327B} for GRB030501.
In all cases the GRB alert was generated and distributed by IBIS, but SPI was also always able to
detect the same event and to confirm in most cases the results obtained with IBIS.
For the  first three bursts the capabilites of SPI were weakened by the telemetry limitations
at the beginning of the mission. The third event was the weak GRB030227 and the GRBs of March and May 2003
were observed at a large offset angle, thus only 15\% to 25\% of SPI's Ge-detectors were irradiated by
the GRB. So currently the demonstration of SPI's full capabilities in the case of a strong
event in the fully-coded FoV still has to take place. An overview on SPI's GRB detection capabilities,
obtained for this first set of GRBs is given by \citet{2003A&A...411L.299V}, which summarises the important 
quantities derived by SPI.

Most of the GRBs were detected by SPI with a S/N between 7 and 16. At this level the GRBs
were located down to error radii of 20' -- 30' (90\% confidence) which are in most cases
overlapping with the one of IBIS. Also the peak flux, fluence and photon indices
are in agreement with the values derived by IBIS. For GRBs observed with SPI operating in full
telemetry mode, spectra could be extracted. In one case (GRB030227) some evidence for a
hard-to-soft spectral evolution was found in the data of both instruments, ISGRI and SPI
\cite{2003ApJ...590L..73M}.
%

\subsection{GRBs observed with the anticoincidence shield ACS}

Since December 2002 the anticoincidence shield (ACS) \cite{2003SPIE.4851.1336V} has been added to the $3^{\rm rd}$ 
interplanetary
network (IPN) of $\gamma$-ray detectors \cite{hurley-grb2003}.
During the first year of the INTEGRAL mission the IPN consisted of Ulysses, Mars Odyssey 2001,
Konus-WIND, HETE-2, RHESSI and INTEGRAL/SPI-ACS. The network had an excellent
configuration, due to the large spacecraft separations between Earth, Mars and Ulysses, which is orbiting
around the sun, out of the ecliptic plane.
The analysis of GRBs detected with the SPI-ACS
during the first eight months (November 2002 - June 2003) of the
INTEGRAL mission is presented below. A more detailed description of these first results, obtained by
SPI-ACS, can be found in \citet{2003A&A...411L.299V}.

As SPI-ACS has no spatial resolution and thus cannot provide the  position of the GRB \cite{2003SPIE.4851.1336V}, the 
GRB nature of a count rate increase
observed by SPI-ACS can only be confirmed by the observation of the same event by another instrument (e.g. Ulysses, 
HETE-2).
These bursts constitute only a subsample of the GRBs detected by SPI-ACS, as different instrumental properties of 
these
 missions (e.g. energy range, sensitivity) do not allow the simultaneous observations/detections of all
GRBs seen by SPI-ACS. Obviously, it is worthwhile to study also bursts which are only visible in the SPI-ACS rates
and not confirmed elsewhere. Probing the very high energies with the unprecedented sensitivity
of the SPI-ACS \cite{2003SPIE.4851.1336V} might open new insights into the burst populations and burst physics.

A sample of possible GRB events, based on the only property measurable by the SPI-ACS, the
veto-countrate lightcurve, is selected for ACS by using only events which exceeds a predefined significance level
above the background (the details of  the selection procedure are described in \citet{2003A&A...411L.299V}). Each 
event is subsequently
checked for solar or particle origin by comparing with events recorded by the X-ray monitor JEM-X and the radiation
monitor IREM of INTEGRAL, or events noted on the GOES web page\footnote{\url{http://www.sec.noaa.gov}}.

\begin{figure}
   \centering
   \includegraphics[width=0.32\textwidth,angle=270]{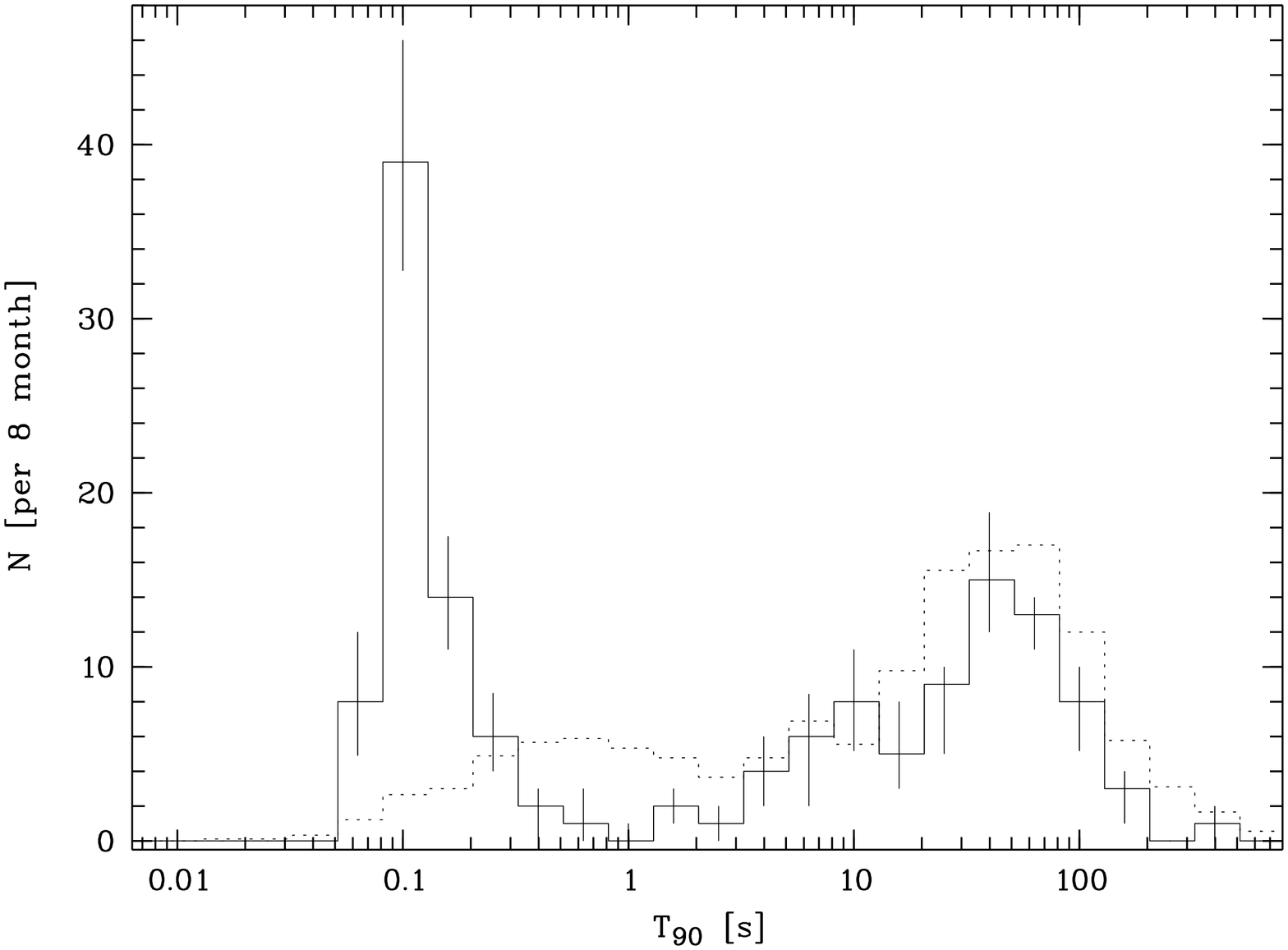}
   \includegraphics[width=0.32\textwidth,angle=270]{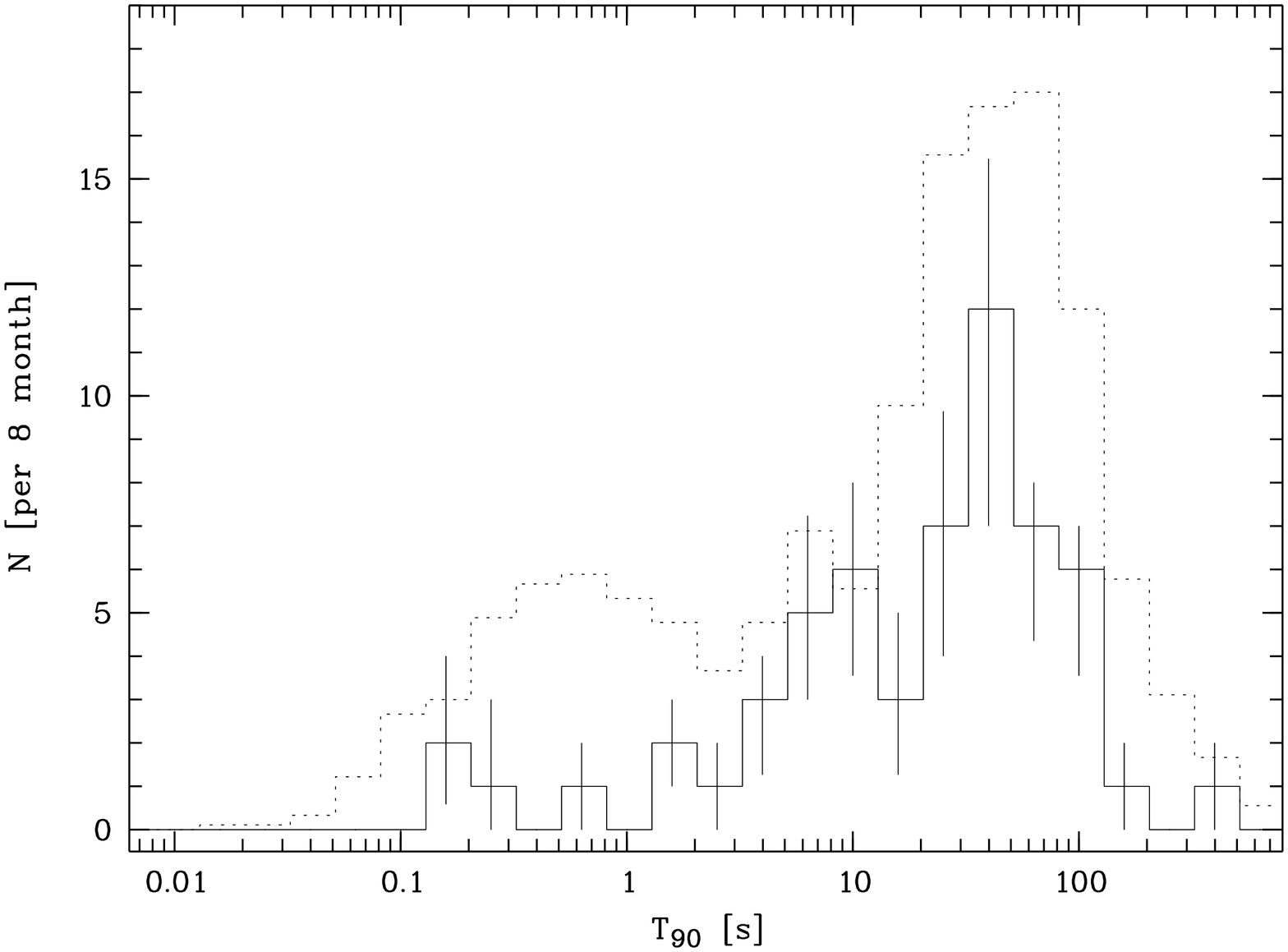}
   \caption { Left: Distribution of T$_{90}$ for all GRB candidates (solid
line) and for 1234 GRBs from the 4th BATSE GRB catalogue 
\cite{1999ApJS..122..465P} (dotted). In order to compare with the SPI-ACS
detections, the BATSE distribution is scaled to the elapsed INTEGRAL
mission time (8 month). Note the very large fraction of short events
compared to BATSE. Right: Same as left plot except that
     here only the confirmed SPI-ACS burst sample (solid line) is
     displayed. While most of the long duration bursts are confirmed
     for comparison an obvious
     lack of short GRBs can be noticed.}
         \label{fig:burstDurationAll}
   \end{figure}

With the selection described above, a total of 145 GRB
candidates were detected during the first 8 months of the mission. 58 of these
have been confirmed by other instruments. Using the elapsed mission
time, we find an approximate rate of GRBs detected by the SPI-ACS of $\sim$290 ($\sim$116
confirmed) per year which is in good agreement with the predictions
given in \cite{2000AIPC..510..722L} prior to the start of the mission. The total
rate is comparable also to BATSE \cite{1999ApJS..122..465P}.
In addition to the number of events, the SPI-ACS overall rate
provides the possibility of deriving the burst duration in the
instrumental observer frame and the variability of the light curve. As
no energy resolution exists, typical burst parameters such as fluence
and peak flux cannot be derived. Only the total integrated counts and
the counts in the burst maximum can be extracted from the light
curve. Fig.~\ref{fig:burstDurationAll} shows the distribution of the measure
for the duration T$_{90}$ (the time interval starting after 5\% and
ending after 95\% of the background subtracted event counts have been
observed) for the sample of SPI-ACS GRBs in comparison to the observed
distribution of 1234 GRBs from the 4$^{\rm th}$ BATSE GRB catalogue
\cite{1999ApJS..122..465P}. Despite the small sample, a bimodality in the
distribution comparable to that found by BATSE is observed. But
two main differences emerge: i) the SPI-ACS sample contains a
significantly higher fraction of short burst candidates and ii) the
maximum of the short distribution is offset towards shorter duration
for the SPI-ACS sample.

The fraction of short ($<$1\,s) duration GRBs is $\sim$0.48 (70/145)
for the SPI-ACS sample compared to 0.20 for BATSE
\cite{1999ApJS..122..465P}. As BATSE was observing a softer energy band
(50-320\,keV) and was therefore more sensitive to X-ray rich (long) GRBs
than SPI-ACS, a larger short/long rate was expected for the ACS
sample.
What is remarkable is the sharpness of the short
distribution around 0.1\,s. Due to the limited time resolution of
50\,ms the short end cannot be sufficiently defined and resolved by
our data.
%
The offset of the maximum for the short events to smaller T$_{90}$
might be due to the different energy bands of SPI-ACS and
BATSE. An apparently shorter duration is measured as it
would be if the bursts would have been observed by BATSE. As T$_{90}$ depends
strongly on the instrumental characteristics and as it is still
unclear how this measure connects to the source frame quantity for a
given burst, the discrepancies are neither surprising nor do they
necessarily  trace different burst populations.
Still, the connection of the short events with real GRBs is not
clear as this population is only marginally observed by other
instruments. While a large
fraction (73\%; 55/75) of the long bursts are confirmed, less than 6\%
(4/70) of the short events were observed by other missions.  This
might be explained at least twofold. On the one hand we might observe
a ``real'' short and very hard GRBs population, which could so far only
be detected with SPI-ACS due to its high sensitivity at very high
energies. As the current IPN members and HETE-2 are generally more
sensitive at lower energies, the detection of a high fraction of un-confirmed short (and
possible hard) events would not be surprising. These bursts should then
have peak energies above 400\,keV. On the other hand, a significant
contribution to these short events  from instrumental effects
and/or cosmic ray events cannot be ruled out. A small contribution
might also arise from soft gamma-ray repeaters (SGRs).  Without localisation
SGR bursts cannot be distinguished from short GRBs within SPI-ACS. The issue of
origin of the short events is certainly of high interest and needs a more
detailed investigation


\begin{theacknowledgments}
The SPI project has been completed under
the responsibility and leadership of CNES. We are grateful
to ASI, CEA, CNES, DLR, ESA, INTA, NASA and OSTC for support.
The SPI/ACS project is supported by the German "Ministerium f\"ur Bildung und Forschung" through
DLR grant 50.OG.9503.0.
\end{theacknowledgments}


\bibliographystyle{aipproc}   

\bibliography{von_kienlin_andreas_0-rev-cor-b}

\IfFileExists{\jobname.bbl}{}
 {\typeout{}
  \typeout{******************************************}
  \typeout{** Please run "bibtex \jobname" to optain}
  \typeout{** the bibliography and then re-run LaTeX}
  \typeout{** twice to fix the references!}
  \typeout{******************************************}
  \typeout{}
 }

\end{document}